\documentclass[%
 aip,
 jmp,%
 amsmath,amssymb,
 preprint,%
]{revtex4-1}

\usepackage{graphicx}% Include figure files
\usepackage{dcolumn}% Align table columns on decimal point
\usepackage{bm}% bold math
\newcommand{\acrofig}{Fig.}
\newcommand{\acroref}{Ref.}

\begin{document}

\preprint{AIP/123-QED}

\title[Current driven domain wall dynamics in ferrimagnetic strips explained by means of a two interacting sublattices model]{Current driven domain wall dynamics in ferrimagnetic strips explained by means of a two interacting sublattices model}
%\thanks{Footnote to title of article.}

\author{Eduardo Mart\'{\i}nez}
\author{V\'{\i}ctor Raposo}%
\affiliation{ 
Dpto. F\'{\i}sica Aplicada, Universidad de Salamanca, 37008 Salamanca, Spain
}%
\author{\'{O}scar Alejos}
\affiliation{%
Dpto. Electricidad y Electr\'{o}nica. Universidad de Valladolid. 47011 Valladolid, Spain%
}%
\email{oscar.alejos@uva.es.}

%\homepage{http://www.Second.institution.edu/~Charlie.Author.}
%\altaffiliation[Also at ]{Physics Department, XYZ University.}

\date{\today}% It is always \today, today,

\begin{abstract}
%The work deals with current-driven domain walls in ferrimagnetic strips. Thin films with perpendicular magnetic anisotropy are considered in two different scenarios. In the first one, the ferrimagnet is grown on a heavy metal, which results in the presence of asymmetric exchange interactions, such as the Dzyaloshinskii-Moriya interaction. Besides, spin currents are generated by means of the spin-Hall effect, when an electric current is injected through the heavy metal. In the second scenario, only spin transfer torques are present when the electric current is injected through the ferrimagnet. 
%The study has been carried out by using a collective coordinates approach, in which the ferrimagnet is considered as a matrix composed by two interacting sublattices, with finite parameters, closely related to those experimentally determined for the matrix components. Results are supported by whole micromagnetic simulations, and has been found to adequately explain recent experimental evidence, and to provide interesting predictions and insights of the current-driven dynamics of domain walls along ferrimagnetic films.
The current-driven domain wall dynamics along ferrimagnetic elements are here theoretically analyzed as a function of temperature by means of micromagnetic simulations and a one dimensional model. Contrarily to conventional effective approaches, our model takes into account the two coupled ferromagnetic sublattices forming the ferrimagnetic element. Although the model is suitable for elements with asymmetric exchange interaction and spin-orbit coupling effects due to adjacent heavy metal layers, we here focus our attention on the case of single-layer ferrimagnetic strips where domain walls adopt achiral Bloch configurations at rest. Such domain walls can be driven by either out-of-plane fields or spin transfer torques upon bulk current injection. Our results indicate that the domain wall velocity is optimized at the angular compensation temperature for both field-driven and current-driven cases. Our advanced models allow us to infer that the precession of the internal domain wall moments is suppressed at such compensation temperature, and they will be useful to interpret state-of-the art experiments on these elements.  
%
%Valid PACS numbers may be entered using the \verb+\pacs{#1}+ command.
\end{abstract}

%\pacs{75.78.Cd,75.78.Fg}% PACS, the Physics and Astronomy
\maketitle

\section{\label{sec:intro}Introduction}
A great effort is being devoted to the finding of optimal systems permitting fast displacement of domain walls (DWs) along racetrack elements.\cite{Parkin:08} As recent experiments demonstrate, DW velocities in the order of $1\frac{\text{km}}{\text{s}}$ can be achieved along ferrimagnetic (FiM) strips,\cite{Kim:17,Caretta:18} with a linear relationship between DW velocities and the magnitude of  applied stimuli.\cite{Kim:17,Caretta:18,Siddiqui:18}

Here we provide a theoretical description of DW dynamics in FiM strips based on an extended collective coordinates model (1DM).\cite{Alejos:18,Martinez:19} Differently from other approaches, based on effective parameters, our model considers such elements as formed by two ferromagnetic sublattices, %with experimentally definite parameters
and coupled by means of an interlattice exchange interaction. Full micromagnetic ($\mu$M) simulations have been performed also to back up those drawn by the 1DM. Importantly, our approaches allow to infer results not achievable from effective models, and to provide insights and interesting predictions of the current-driven dynamics of DWs along FiM films.

\begin{figure}[ht]
\centering
  \begin{tabular}{@{}cc@{}}
    (a)\includegraphics[scale=0.9]{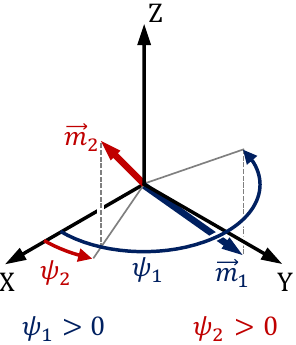} &
    (b)\includegraphics[scale=0.9]{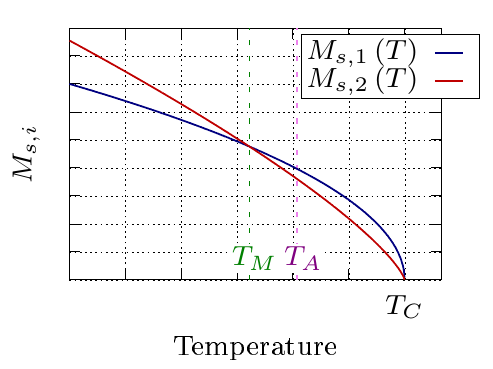}\\
    \multicolumn{2}{c}{(c)\includegraphics[scale=1]{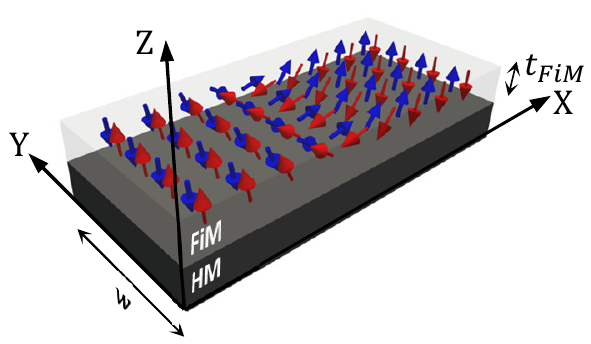}}\\
    \multicolumn{2}{c}{(d)\includegraphics[scale=1]{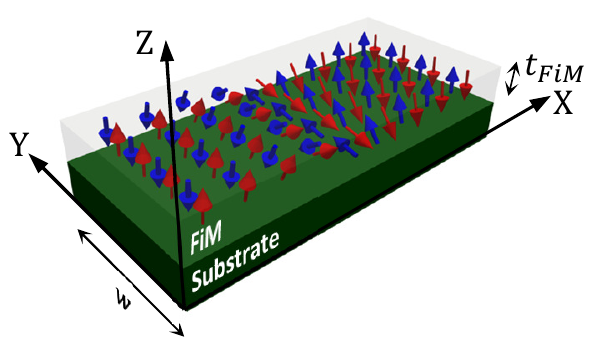}}
  \end{tabular}
  \caption{Two sublattices constitute the FiM: (a) magnetizations are represented by the unit vectors $\vec{m}_1$ and $\vec{m}_2$, with in-plane orientation angles $\psi_1$ and $\psi_2$, respectively, (b) temperature dependence of the magnetization of each sublattice, (c) magnetic DW of N{\'e}el type, and (d) magnetic DW of Bloch type amidst two domains oriented out of plane (the strip width $w$ is here shown).}
  \label{Fig:01}
\end{figure}

\acrofig\ref{Fig:01}.(a) schematizes the local orientation of magnetic moments in the ferrimagnet. $\vec{m}_i \left(i=1,2\right)$ represent the orientations of the respective magnetic moments of each ferromagnetic sublattice. The magnetization of each sublattice is temperature dependent, %in the typical way, that is, 
so that magnetization of each sublattice vanishes at Curie temperature ($T_C$), with a magnetization compensation temperature $T_M$, as it is shown in \acrofig\ref{Fig:01}.(b). The temperature dependence %of each magnetization 
can be described by the analytical functions: $M_{s,i}\left(T\right)=M_{s,i}^0\left(1-\frac{T}{T_C}\right)^{a_i}$, $M_{s,i}^0$ being the respective magnetizations at zero temperature, and $a_i$ being dependent on the sublattice components. 

The model can be applied to two different architectures. As a first architecture (\acrofig\ref{Fig:01}.(c)), a FiM strip on top of a heavy metal (HM) can be considered. The FiM/HM interface promotes interfacial asymmetric exchange, resulting in N{\'e}el type DWs and current driven domain wall motion (CDDWM) due to spin orbit torques (SOT), with rigid DWs. At the angular momentum compensation temperature ($T_A$), differing from $T_M$ due to the distinct Land{\'e} factors $g_i$ for each sublattice, DW magnetic moments keep aligned with the current, leading to a linear increase of DW velocities. Thus, DW velocities are maximized at $T_A$. This first architecture has already been adequately discussed from both the experimental\cite{Caretta:18} and theoretical\cite{Caretta:18,Martinez:19} points of view, in particular, by using the model to be here recalled\cite{Martinez:19}. In the second architecture (\acrofig\ref{Fig:01}.(d)), the FiM does not lie on a HM, and so interfacial asymmetric exchange vanishes. CDDWM is dominated by the spin transfer torques (STT), and DW precessional regimes emerge, due to reduced magnetostic interactions, resulting in DW velocities proportional to current magnitudes. Again, DW velocities have been found to maximize at $T_A$, when precession freezes, leading to a CDDWM characterized by rigid DWs, what is to be shown along this text.

\section{\label{sec:1DM} Two-sublattice model of ferrimagnets}
The description of the DW dynamics by means of a 1DM starts from the application of variational principles to the $\mu$M equation, i.e, the Landau-Lifshitz-Gilbert (LLG) equation.\cite{Haazen:13,Zhang:04} This procedure is then augmented to study the magnetization dynamics in FiMs by posing two coupled LLG equations, that is, a two-sublattice model (TSLM). Details on the derivation of the 1DM equations for the TSLM are given in \acroref\onlinecite{Martinez:19}, so here we will only recall the required model parameters.

Within the model, the respective Gilbert constants of each sublattice are represented by the values $\alpha_i$. The effective fields are the sum of the external field, the demagnetizing (magnetostatic) fields, the anisotropy fields, the isotropic exchange fields and the asymmetric exchange fields. The external field have components $\left(B_x,B_y,B_z\right)$. The demagnetizing term possesses out-of-plane and in-plane components, given by the effective anisotropy constants $K_{eff,i}$ and $K_{sh,i}$. The asymmetric exchange provides a chiral character to some magnetic textures, whereas the isotropic one can be reduced on first approach to the sum of an intra-sublattice exchange field, given by the exchange stiffness $A_i$, and an inter-sublattice interaction due to the misalignment of both sublattices. The latter is accounted for by a parameter $B_{12}>0$ ($<0$), which promotes the antiparallel (parallel) alignment of the sublattices. Finally, LLG equations also include the torques due to spin polarized currents, i.e., the STT\cite{Haazen:13} and the SOT\cite{Zhang:04}. Here, we focus our attention on the STT, consisting of adiabatic interactions and their non-adiabatic counterparts. The adiabatic interactions are defined by values $u_i$, proportional to the electric density current $J_x$ flowing along the element, and calculated as $u_i=\frac{1}{2}\frac{g_i\mu_BP}{eM_{s,i}}J_x$, with $\mu_B$ being Bohr's magneton, $e$ the electron charge, and $P$ the degree of polarization of the spin current. The non-adiabatic interactions are proportional to the adiabatic ones by factors $\beta_i$.

The derivation of the 1DM requires the DW profile to be described in terms of the DW position $q$, width $\Delta$ and transition type $Q$. In the TSLM, the DW is considered to be composed of two transitions, one for each sublattice, which share the same $q$, and the same $\Delta$ (see \acrofig\ref{Fig:01}.(c) and (d)), but $Q_i=\pm 1$ establishes the transition type for each sublattice. $Q_i=+1\left(-1\right)$ means up-down (down-up) transition. Due to the antiferro coupling between sublattices, it follows that $Q_1=-Q_2$.

\section{Results and discussion}

When FiMs, such as GdFeCo or Mn$_4$N, are grown on top of certain substrates, the absence of interfacial asymmetric exchange\cite{Kim:17,Gushi:19} results in the formation of achiral DWs. The orientation of DW internal moments at rest is then dependent on purely geometrical aspects. In particular, for thin strips sufficiently wide, magnetostatic interactions determine the formation of Bloch-type walls. %
%Walker field can be estimated as 2\pi\alpha(MsU-MsL)*(Nx-Ny)\cite{Mougin:07}
%Hk=2.2495e+04
%Bw=3.5523mT
%Walker current can be estimated as \gamma0\Delta(MsU-MsL)*(Nx-Ny)/2\cite{Thiaville:04,Thiaville:05}
Importantly, due to the low net magnetization of FiMs as compared with ferromagnets, the magnetostatic interactions are rather low. If some parallelism between ferro- and ferrimagnets is made, Walker breakdown in FiMs is then expected to occur for rather low applied fields\cite{Mougin:07} or currents\cite{Thiaville:04,Thiaville:05} in the temperature range around $T_M$. Consequently, the DW dynamics for moderate fields or currents is ruled by the precession of DW magnetic moments.

The case of the field-driven DW dynamics in ferrimagnetic GdFeCo alloys can be recalled at this point. This has been the subject of recent experimental work,\cite{Kim:17} where fast field-driven antiferromagnetic spin dynamics is realized in FiMs at $T_A$. This behavior has been found to be reproducible with the TSLM. Our simulations have been carried out with a set of parameters similar to those considered in previous works,\cite{Caretta:18, Martinez:19} but adapted as to take into account the absence of interfacial asymmetric exchange and SOTs. The parameters are: $A_i=70\frac{\text{pJ}}{\text{m}}$, $K_{eff,i}\approx K_{u,i}= 1.4\frac{\text{MJ}}{\text{m}^3}$, $K_{u,i}$ being the magnetic uniaxial anisotropy constant of the FiM sublattices. With these parameters, DW width is $\Delta\approx 6\text{nm}$. Besides, $\alpha_i = 0.02$.  Due to the low net magnetization in the temperature range of interest, $K_{sh,i}\approx 0$. The antiferromagnetic coupling is accounted for by the parameter $B_{12} = 9\frac{\text{MJ}}{\text{m}^3}$.\cite{Ma:16} The gyromagnetic ratios ($\gamma_i = \frac{g_i\mu_B}{\hbar}$) are different due to distinct Land{\'e} factors: $g_1 = 2.2$ and $g_2 = 2.0$.\cite{Kim:17} The Curie temperature is set to $T_C = 450\text{K}$, and $M_{s,1}^0 = 1.4\frac{\text{MA}}{\text{m}}$ and $M_{s,2}^0 = 1.71\frac{\text{MA}}{\text{m}}$, with $a_1 = 0.5$ and $a_2 = 0.76$. According to these values, $T_M \approx 241.5\text{K}$, and $T_A\approx 305\text{K}$. The dimensions of the FiM strips are $w\times t_{FiM}=512\text{nm}\times 6\text{nm}$.
 
%\begin{figure}[ht]
%\includegraphics[scale=1]{VvsHz}
%\caption{Terminal velocity as a function of the out-of-plane applied field $B_z$ at different temperatures. Dots and continuous lines correspond respectively to full $\mu$M simulations and the 1DM results.}
%\label{Fig:02}
%\end{figure}

\begin{figure*}[ht]
\centering
  \begin{tabular}{@{}lll@{}}
    (a)\includegraphics[scale=0.6]{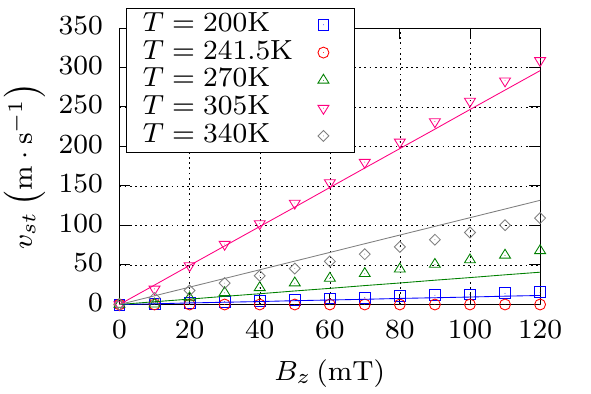} &
    (d)\includegraphics[scale=0.6]{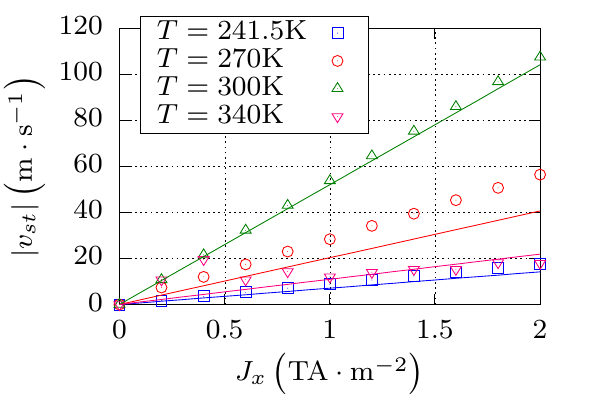}  &
    (g)\includegraphics[scale=0.6]{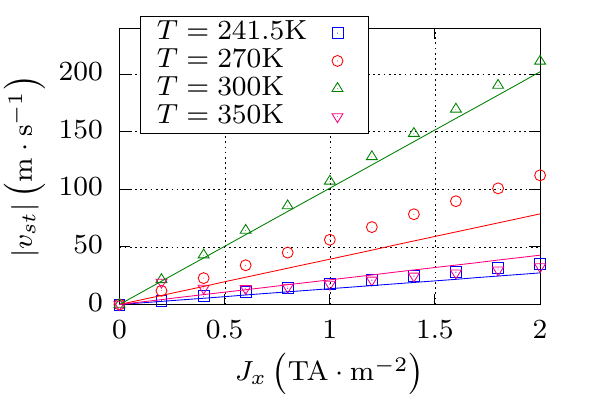} \\
    (b)\includegraphics[scale=0.6]{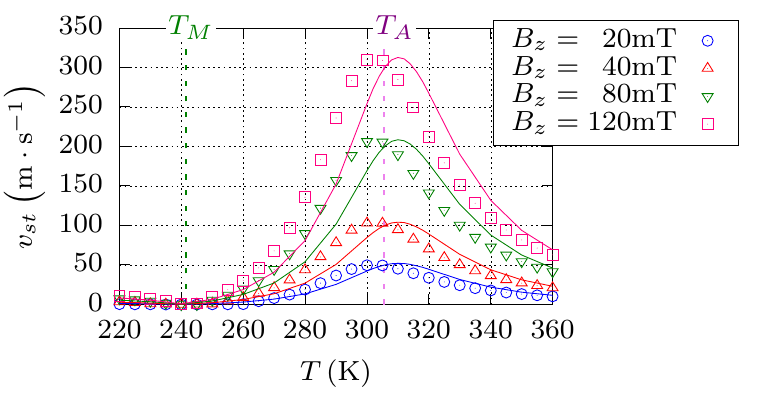} &
    (e)\includegraphics[scale=0.6]{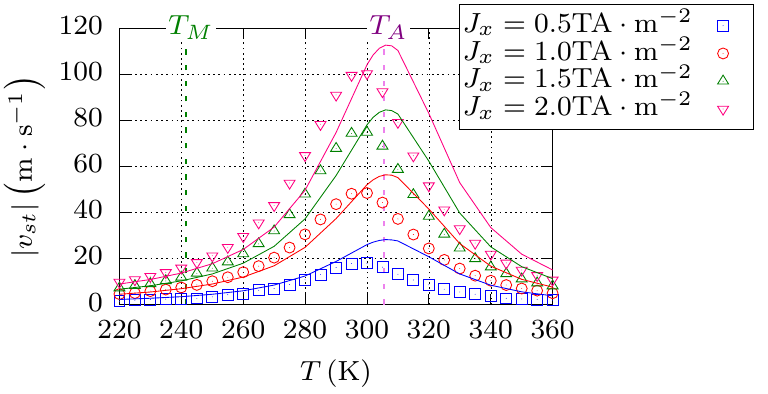} &
    (h)\includegraphics[scale=0.6]{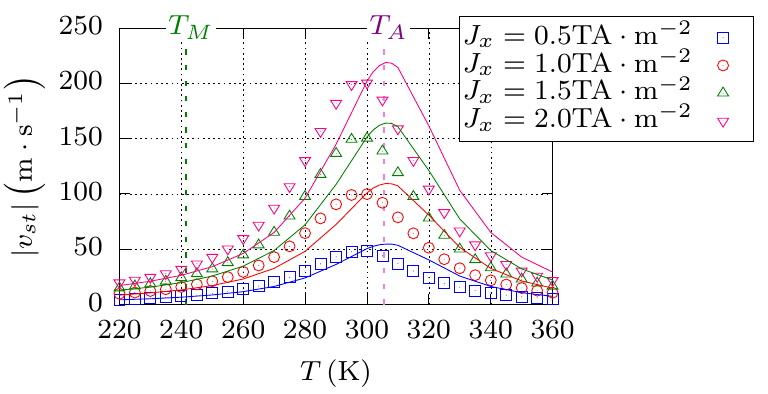}\\
    (c)\includegraphics[scale=0.6]{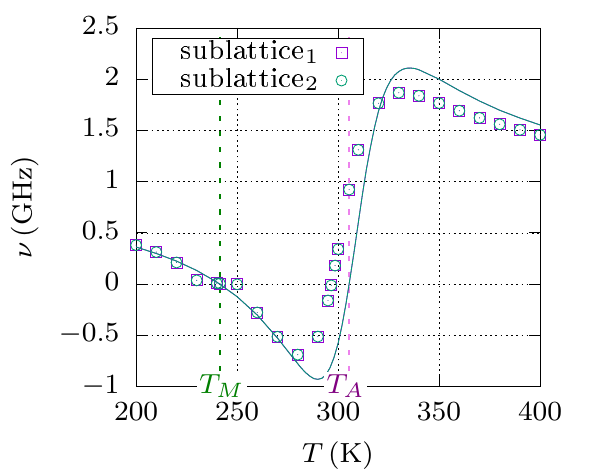} &
    (f)\includegraphics[scale=0.6]{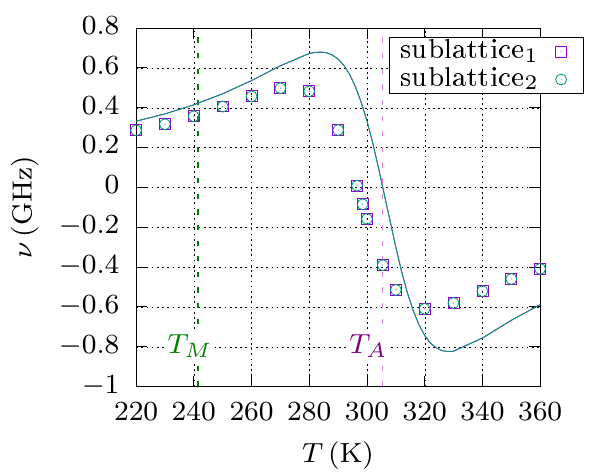} &
    (i)\includegraphics[scale=0.6]{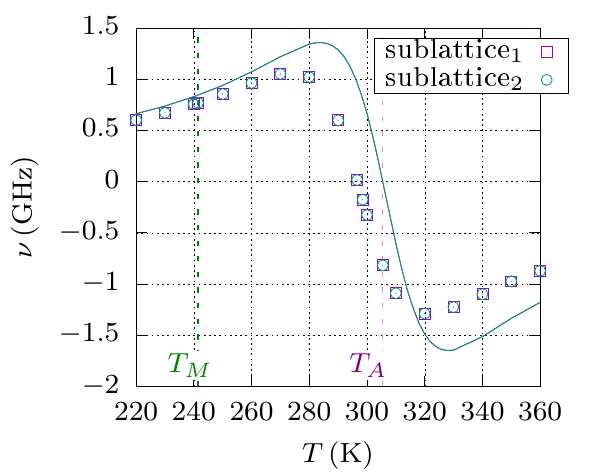}
  \end{tabular}
  \caption{Field-driven an current-driven dynamics in a FiM strip: (a) terminal velocity as a function of $B_z$ with temperature as a parameter, (b) terminal velocity with $B_z$ as a parameter and (c) precessional frecuencies of DWs for $B_z=40\text{mT}$ as functions of temperature, (d) and (g) terminal velocity as a function of $J_x$ with temperature as a parameter, (e) and (h) terminal velocity with $J_x$ as a parameter and (f) and (i) precessional frecuency of DWs for $J_x=1\frac{\text{TA}}{\text{m}^2}$ as functions of temperature. $\beta_i=\alpha_i$ for (d), (e) and (f), whereas $\beta_i=2\alpha_i$ for (g), (h) and (i). Dots and continuous lines correspond respectively to full $\mu$M simulations and the 1DM results.}
  \label{Fig:03}
\end{figure*}

\acrofig\ref{Fig:03}.(a) presents the dependence of the DW terminal velocity, computed as $v_{st}=\frac{q\left(\Delta t\right)-q\left(0\right)}{\Delta t}$, with $\Delta t=2\text{ns}$, on the out-of-plane applied field $B_z$ at different temperatures. In agreement with experiments,\cite{Kim:17} $v_{st}$ increase linearly with $B_z$, and the slope reaches a maximun at $T_A$. This fact is made clear in \acrofig\ref{Fig:03}.(b) where terminal velocity is represented as a function of temperature with $B_z$ as a parameter. In all shown cases, no dynamics occurs at $T_M$ since the net magnetization vanishes, whereas the highest speeds are found close to $T_A$. The clue for this behavior can be found in DW precession, represented as a function of temperature in \acrofig\ref{Fig:03}.(c). Precession frequencies are obtained as $\nu=\frac{\dot\psi_i\left(\Delta t\right)}{2\pi}$ ($i=1,2$), since $\dot\psi_1\left(\Delta t\right)\approx\dot\psi_2\left(\Delta t\right)$. The results demonstrate that during the dynamics, DW magnetic moments precess except at temperatures around $T_M$ and $T_A$, where precession freezes and the orientation of DW magnetic moments during the whole dynamics holds.

Previous field-driven analysis serves as a starting point to also understand the CDDWM in these elements. This dynamics is purely governed by STT because DWs move contrary to the current direction.\cite{Gushi:19} \acrofig\ref{Fig:03}.(d) and (g) present the dependence of the absolute terminal velocity as a function of the current $J_x$ with the temperature as a parameter. The polarization has been set to $P=0.7$, and the non-adiabatic transfer torque parameters have been chosen as (d) $\beta_i=\alpha_i$ (also for figures (e) and (f)), and (g) $\beta_i=2\alpha_i$ (also for figures (h) and (i)). Differently from the results obtained in the field-driven case, the CDDWM at $T_M$ is not null, since the STT pushes the transitions in each sublattice in the same direction (and not in opposite directions as it occurs in the field-driven case). However, the maximum slope is again found at $T_A$, when the precessional frequency vanishes.

%\begin{figure*}[ht]
%\centering
%  \begin{tabular}{@{}ll@{}}
%    (a)\includegraphics[scale=1]{VvsT_xi0p02} & (c)\includegraphics[scale=1]{VvsT_xi0p04}\\
%    (b)\includegraphics[scale=1]{PhivsT_Jx_xi0p02} & (d)\includegraphics[scale=1]{PhivsT_Jx_xi0p04}
%  \end{tabular}
%  \caption{Terminal velocity and precessional frecuencies of DWs in a FiM strip as a function of temperature with $J_x$ as a parameter. The cases shown are: (a) terminal velocity for $\beta_i=\alpha_i$, (b) precessional frequency for $\beta_i=\alpha_i$ (c) terminal velocity for $\beta_i=2\alpha_i$, and (d) precessional frequency for $\beta_i=2\alpha_i$. % Dots and continuous lines correspond respectively to full $\mu$M simulations and the 1DM results.}
%  \label{Fig:05}
%\end{figure*}

%The plot of the absolute terminal velocity as a function of temperature shown in \acrofig\ref{Fig:05}.(a) and (c) confirms that the fastest dynamics occurs again at $T_A$. Besides, the precessional frequencies are plotted in \acrofig\ref{Fig:05}.(b) and (d). It can be found again that precession freezes around $T_A$.

\begin{figure*}[ht]
\includegraphics[scale=1]{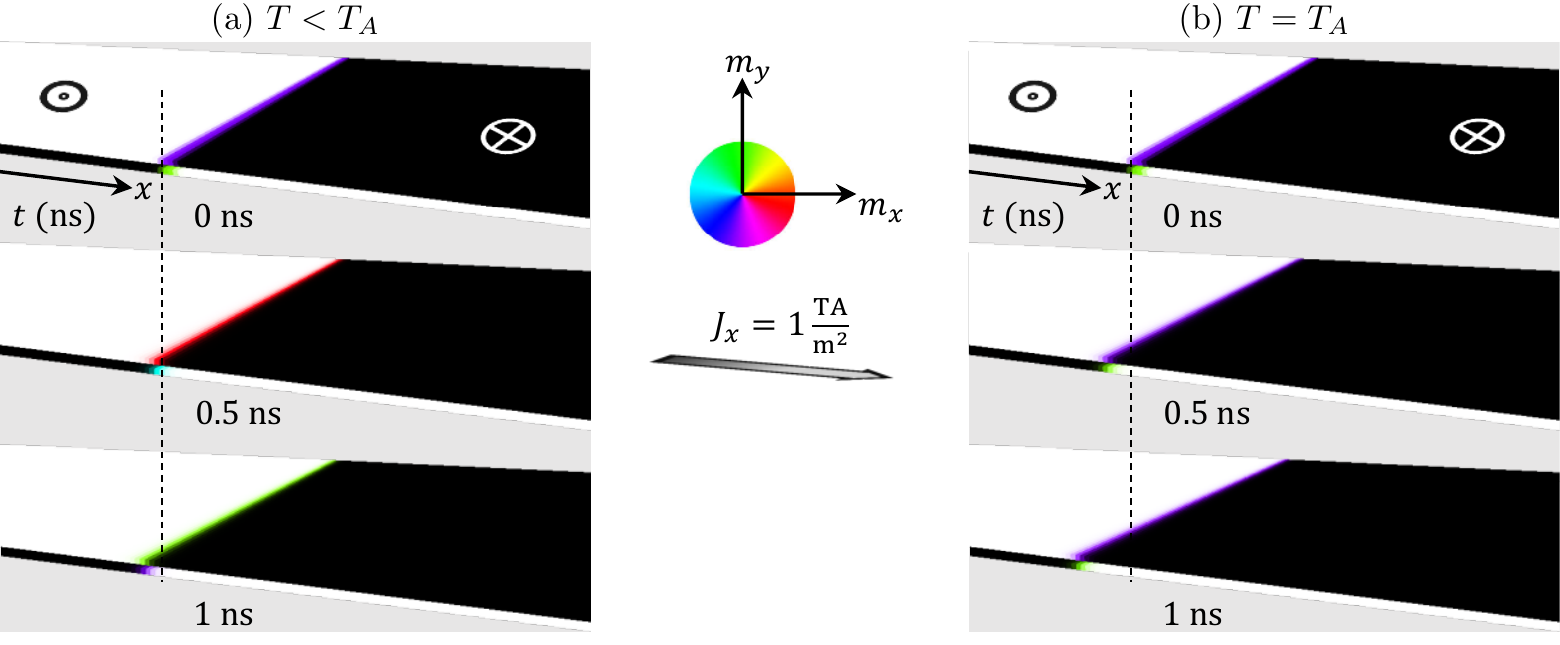}
\caption{Snapshots of the CDDWM in a FiM strip with $\beta_i=\alpha_i$ at (a) $T<T_A$, and (b) $T=T_A$.}
\label{Fig:06}
\end{figure*}

To show in more detail this behavior, \acrofig\ref{Fig:06} presents the snapshots of the CDDWM at two representative temperatures, for the case $\beta_i=\alpha_i$. The two sublattices composing the FiM are presented superposed, as to simplify the view, so one sublattice is on top of the other. The images in (a) correspond to the dynamics at $T<T_A$. In this case, the DW internal moments precess, and a turn of approximately $180\textsuperscript{o}$ takes place within the $1\text{ns}$-interval passing from the image on top to the bottom image. However, no precession takes place at $T=T_A$, as shown in (b). The distinct distances run by the DWs can be also compared.

Differently from the behavior of magnetic moments in pure ferromagnets, where STT compensates damping when $\beta_i=\alpha_i$, FiMs seem to present these precessing magnetic moments even in this case. Such precession would be associated with the torque due to the coupling between the two sublattices and freezes at $T_A$, a result that would not be in any case explainable by means of effective models.

\section{Conclusions}

The aim of this work has been first to highlight the capacities of the TSLM and, particularly, the 1DM based on it, to reproduce recent experimental work on DW dynamics in FiMs. Differently from previous approaches, the TSLM does not require the use of effective parameters, but experimentally determined ones, which allows providing insightful details about the dynamics. 

The work has been devoted to FiMs structured so that DWs adopt achiral Bloch configurations at rest, and the main conclusions of this work are as following. The DW dynamics in FiMs is characterized by DW precession, which freezes at $T_A$. Because of that, the DW velocity at $T_A$ is enhanced, both for the field- and for the current-driven cases. Our results are also in good qualitative agreement with recent experimental observations: \acroref\onlinecite{Kim:17} for the field-driven case, and \acroref\onlinecite{Gushi:19} for the current-driven one. Finally, the physical origin or the fundamental reasons behind these observations can only be achieved by adopting models which consider the independent but antiferromagnetically coupled nature of the two sublattices forming the FiM. Therefore, our models will be useful to understand state-of-the-art experiments and also to develop and optimize future DW-based devices.

%Additionally, the work essays new interesting and promising results about the CDDWM due to STT.
%, where, for example, large $\beta_i$ as compared with $\alpha_i$ may give rise to extraordinarily fast DWs in these elements, a fact that is being under study nowadays.

\section{Acknowledgement}

This work was partially supported by Project No. MAT2017-87072-C4-1-P from the (Ministerio de Econom{\'\i}a y Competitividad) Spanish Government and Project No. SA299P18 from the (Consejer{\'\i}a de Educaci{\'o}n) of Junta de Castilla y Le{\'o}n.

\section{Bibliography}
%\nocite{*}
\bibliography{\jobname}% Produces the bibliography via BibTeX.
%\bibliographystyle{elsarticle-harv}
%\biboptions{authoryear}

\end{document}